\documentclass{article}
\usepackage{spconf,amsmath,graphicx}

\usepackage{lineno,hyperref}

\usepackage{epsfig,subfigure}
\usepackage{amsmath}
\usepackage{amssymb}
\usepackage{graphicx}
\usepackage{setspace}
\usepackage{multirow}
\usepackage[table]{xcolor}
\usepackage{amsmath,array,graphicx}
\modulolinenumbers[5]
\usepackage[german]{babel}
\usepackage{scalerel}
\usepackage{ upgreek }

\usepackage{float}
\usepackage{stfloats}

\usepackage{textcomp}

\usepackage{graphicx}
\usepackage{booktabs}

\usepackage{array,tabularx}
\usepackage{gensymb}

\usepackage[
		     sort&compress]{natbib}
		     		     
\usepackage[table]{xcolor}
\usepackage{makecell}

\usepackage[symbol]{footmisc}

\usepackage{enumitem}

\makeatletter

\renewcommand\paragraph{\@startsection{paragraph}{4}{\z@}%
                                     {-3.25ex\@plus -1ex \@minus -.2ex}%
                                     {1.5ex \@plus .2ex}%
                                     {\normalfont\normalsize\bfseries}}
                                     
\newcommand{\norm}[1]{\left\lVert#1\right\rVert}

\usepackage{textgreek}

\DeclareMathOperator*{\argmax}{arg\,max}
\DeclareMathOperator*{\argmin}{arg\,min}
\usepackage{siunitx}
\definecolor{LightGray}{rgb}{0.92,0.92,0.92}

\let\footnotemark

\begin{document}

\title{Probabilistic combination of eigenlungs-based classifiers for COVID-19 diagnosis in chest CT images}


\def\@name{ \emph{Juan E. Arco$^{1,*}$\thanks{\textsuperscript{*} Corresponding author: jearco@ugr.es}},  \emph{Andr\'es Ortiz$^{2}$}, \emph{Javier Ram\'irez$^{1}$}, \emph{Francisco J. Mart\'inez-Murcia$^{2}$}, \\  \emph{Yu-Dong Zhang$^{3}$}, \emph{Jordi Broncano$^{4}$}, \emph{M. \'Alvaro Berb\'is $^{4}$}, \emph{Javier Royuela-del-Val$^{4}$}, \\ \emph{Antonio Luna$^{5}$},  \emph{Juan M. G\'orriz$^{1}$}}


\address{\normalsize $^{1}$ Department of Signal Theory, Networking and Communications, Universidad de Granada\\ 
\normalsize $^{2}$ Department of Signal Theory, Networking and Communications, Universidad de Malaga \\
\normalsize $^{3}$ School of Informatics, University of Leicester, Leicester, LE1 7RH, Leicestershire, UK \\
\normalsize $^{4}$ Department of Radiology, Hospital San Juan de Dios, HT M\'{e}dica \\
\normalsize $^{5}$ Department of Radiology, Cl\'inica Las Nieves, HT M\'{e}dica \\
}

\maketitle

\begin{abstract}
The outbreak of the COVID-19 (Coronavirus disease 2019) pandemic has changed the world. According to the World Health Organization (WHO), there have been more than 100 million confirmed cases of COVID-19, including more than 2.4 million deaths. It is extremely important the early detection of the disease, and the use of medical imaging such as chest X-ray (CXR) and chest Computed Tomography (CCT) have proved to be an excellent solution. However, this process requires clinicians to do it within a manual and time-consuming task, which is not ideal when trying to speed up the diagnosis. In this work, we propose an ensemble classifier based on probabilistic Support Vector Machine (SVM) in order to identify pneumonia patterns while providing information about the reliability of the classification. Specifically, each CCT scan is divided into cubic patches and features contained in each one of them are extracted by applying kernel PCA. The use of base classifiers within an ensemble allows our system to identify the pneumonia patterns regardless of their size or location. Decisions of each individual patch are then combined into a global one according to the reliability of each individual classification: the lower the uncertainty, the higher the contribution. Performance is evaluated in a real scenario, yielding an accuracy of 97.86\%. The excellent results obtained  and the simplicity of the system (use of deep learning in CCT images would result in a huge computational cost) evidence the applicability of our proposal in a real-world environment. 

\footnote{2022 IEEE.  Personal use of this material is permitted.  Permission from IEEE must be obtained for all other uses, in any current or future media, including reprinting/republishing this material for advertising or promotional purposes, creating new collective works, for resale or redistribution to servers or lists, or reuse of any copyrighted component of this work in other works.}

\end{abstract}


\section{Introduction}
\label{sec:intro}

The COVID-19 (Coronavirus disease 2019) pandemic has had a dramatic effect in global health. According to the World Health Organization (WHO), there have been more than 100 million confirmed cases of COVID-19, including more than 2.4 million deaths \citep{who1}. The development and rapid distribution of vaccines is crucial for fighting against this disease. However, it is also extremely important the early detection of the disease, especially because of many people are contagious in the pre-symptomatic period \citep{who2}. Reverse-transcription polymerase chain reaction (RT-PCR) is usually employed for the detection of COVID-19 \citep{pcr1,pcr2}. This method relies on detecting viral RNA fragments from nasopharyngeal swab. Despite RT-PCR is the gold standard, it has some limitations such as the large time needed until results are obtained and the relatively low sensitivity. In fact, some patients have typical clinical symptoms but still developing a false negative PCR \citep{pcr3}, increasing the risk of community transmission and delaying the medical treatment. An alternative solution is the use of medical imaging such as chest X-ray (CXR) \citep{x-ray} and chest Computed Tomography (CCT) \citep{ct}. The low cost of X-ray imaging has popularized its use as a diagnostic tool for pneumonia. However, CCT collects high-quality-3D volumetric images that outperforms the 2D images obtained by CXR, making CCT an excellent tool for the identification of the ground glass opacities (GGO) typically present in pneumonia caused by COVID-19 \citep{HANI2020263,opacities2}

Despite the high spatial resolution of CCT images \citep{ct1}, pneumonia diagnosis is not a straightforward task. Success depends on factors such as the expertise of the radiologist \citep{chandra2020} or the overlapping between symptoms associated with pneumonia and abnormal lung conditions \citep{maduskar2016}. This leads to a manual, time-consuming process that may delay diagnosis and the election of a correct treatment. For this reason, the use of methods based on machine/deep learning can play a decisive role in the automation of this process. Previous studies have employed machine learning methods for the automatic detection of pneumonia \citep{SOUSA20132579,KHAN2021106960,chandra2020,elaziz2020}. Some of them have been used for delimitating lungs boundaries as an initial step of the classification system \citep{guan2020,yang2018,vajda2018}, whereas others have been successfully employed in the classification stage \citep{porcel2008,zhang2020,ma2015,ajin2017,yahyaoui2018,chandra2020}. The emergence of deep neural networks has revolutionized the automatic classification of medical images, including pneumonia detection \citep{varshni2019,kermany2018,mittal2020}. \cite{rajpurkar2017} proposed a 121-layer convolutional neural network (CNN) that provided a relatively low accuracy (76.8\%), but in a multiclass classification between 14 different pathologies. \cite{ELKORANY2021166405} recently presented COVIDetection-Net, a deep learning model to detect and classify COVID-19 and other types of pneumonia from CXR images. Their system yielded 94.44\% in a 4-class classification, leading to up to 100\% when distinguishing between controls and pneumonia patients. \cite{EZZAT2021106742} proposed a CNN architecture based on the gravitational search algorithm (GSA) to diagnose COVID-19 disease. Specifically, this approach adopted a transfer learning method using a pre-trained DenseNET121 whose hyperparameters were optimized by GSA to maximize performance, yielding an accuracy of 98.38\%.


Complex machine learning tools and deep learning models have demonstrated a good performance in the identification of pneumonia patterns. However, complexity can have a dramatic effect in performance when information to be extracted is high. In fact, most of these models are based on CXR images since they are two-dimensional, alleviating considerably the computational cost \citep{measurement2020,mittal2020,chandra2020}. When applying to CCT images, one solution is to select only one slice, so that input images have two dimensions as CXR. \cite{WANG2021131} successfully employed this alternative within a transfer learning setting and discriminant correlation analysis. \cite{Zhang2021COVID19DV} also used it in combination with DenseNet, leading to an accuracy of 96\%. However, this extreme dimensionality reduction leads to a vast loss in the spatial information that CCT images provide. In fact, this is essentially the advantage of CCR over CXR images: higher resolution and volumetric data. Thus, eliminating the third dimension can drastically mitigate the potential of CCT images. Another important issue is the way the target slice is selected from the CCT scan, and this is not a trivial task. One possibility is to choose the slice displaying the largest number of lesions and size. However, this process requires clinicians to do it manually, which is not ideal when trying to automatize the diagnosis. The simplest solution is to employ the whole CCT image and not selecting only one slice, but this entails a high computational cost. This workload would be even worse when using deep learning and three-dimensional convolutions for feature extraction. Besides, training a deep learning model requires a considerable amount of data, especially when the number of features is high like in 3D images. Thus, it is necessary to find a solution that leads to a high classification performance while mitigating the computational load associated with the processing of volumetric data. 

\begin{figure*}[t]
\centering
\includegraphics[width=0.6\textwidth]{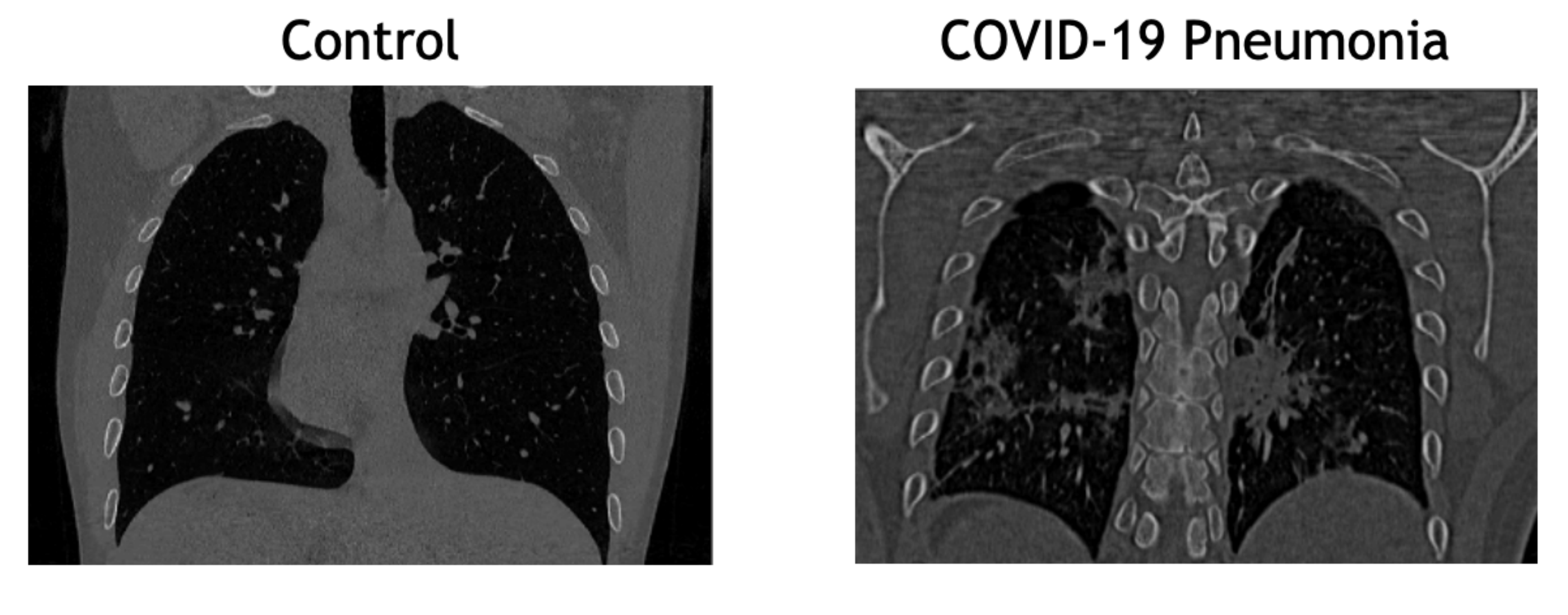}
\caption{Slices of two CCT images of a control (left) and a pneumonia caused by COVID-19 (right). Note some clear artifacts in COVID-19 image.}
\label{fig:figurauno}
\end{figure*}

The collapse of the health system worldwide has manifested the importance to find a rapid, simple and accurate method for detecting pneumonia associated with COVID-19. In this work, we employ an ensemble classifier based on probabilistic SVM in order to identify pneumonia patterns while providing information about the reliability of the classification. In particular, each CCT scan is divided into a number of cubic patches. Features contained in each one of them are extracted by applying kernel PCA, and the most informative components are then entered into an RBF-SVM classifier. The use of base classifiers within an ensemble allows our system to identify the pneumonia patterns regardless of their size or location. Decisions of each individual patch are then combined into a global one according to the reliability of each individual classification: the lower the uncertainty, the higher the contribution, and vice versa. Performance is evaluated in a real scenario where distinguishing between controls \textit{vs} pneumonia patients. The main contributions of our work can be summarized as follows:

\begin{itemize}
\item{A novel and accurate tool for the detection of pneumonia associated with COVID-19 from CCT images.}
\item{Probabilistic classifiers quantify the reliability of their predictions.}
\item{This information is then used to weigh the contribution of each classifier to the final decision.}
\item{The ensemble scheme allows the identification of pneumonia patterns regardless of their position and extension.}
\item{This system alleviates the computational cost of 3D images processing while yielding a high performance.}
\end{itemize}

\section{Materials}
\label{sec:materials}

\subsection{Dataset}
\label{subsec:dataset}
The dataset employed in this work was provided by HT M\'{e}dica, a company specialized in radiology that offers innovative solutions for image diagnosis. The dataset comprises 513 CCT images, including 100 control patients and 413 characterized as depicting pneumonia associated with COVID-19.  All images were obtained as part of patient's routines clinical care during the first wave of the COVID-19 pandemic in Spain (March to June 2020). Data were anonymized before being used in this study following the requirements stated by medical ethics committees. Figure \ref{fig:figurauno} shows a slice of a CCT scan from a control (CL) and a patient suffering from pneumonia (PNEU).

\subsection{Image preprocessing}
\label{subsec:prepro}
\begin{figure*}
\centering
\includegraphics[width=0.6\textwidth] {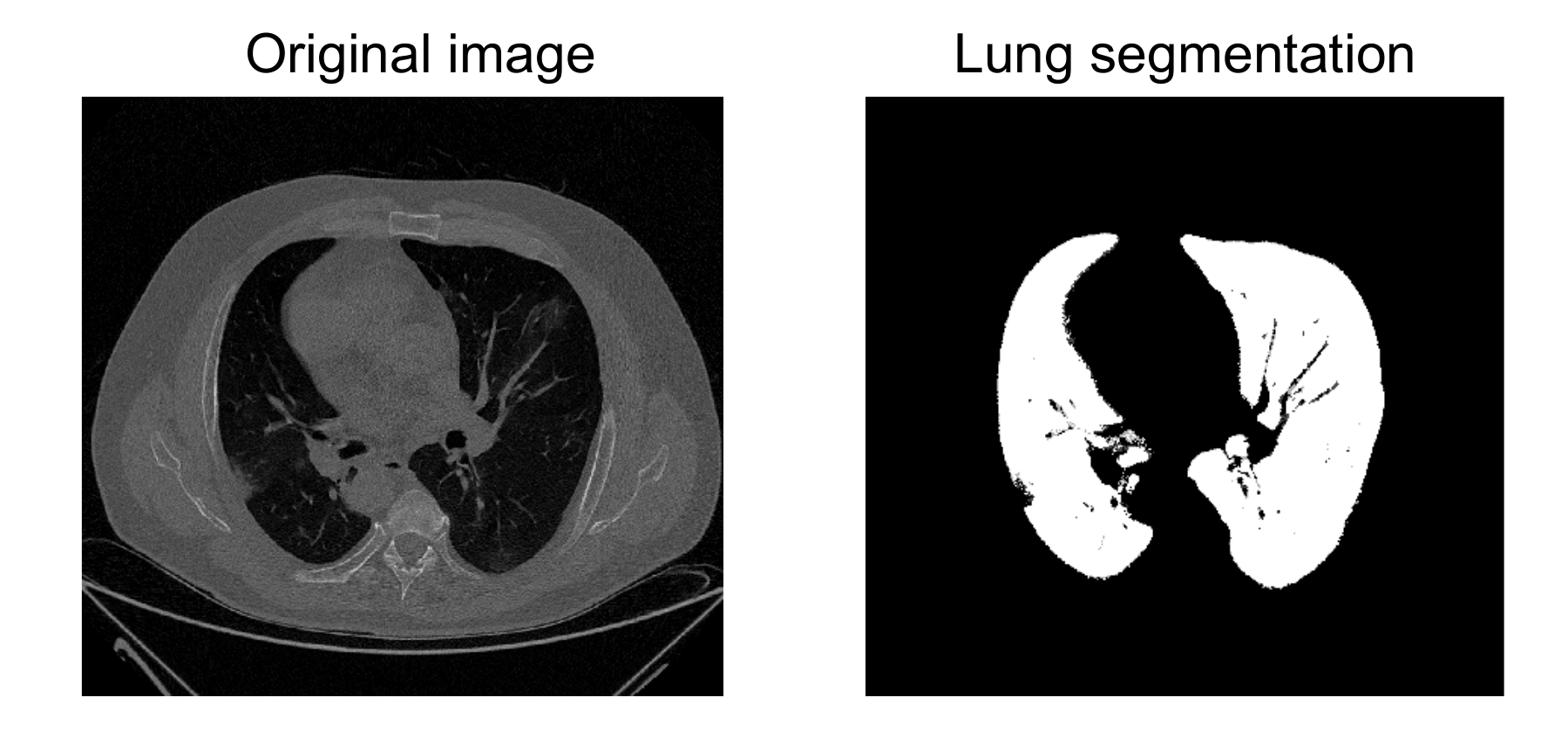}
\caption{Original image (left) and the one obtained after lung segmentation (right).}
\label{fig:lung_segmentation}
\end{figure*}

When working with medical images, it is crucial to apply preprocessing in order to improve the subsequent classification performance. Therefore, this operation must adapt images to the requirements of the classification framework. Given the high computational and memory requirements of CCT volumes, we downsampled the input images to obtain a final map of 128x128x128. We also performed an automatic lung segmentation in order to separate voxels corresponding to lung tissues from those of the surrounding anatomy. To do so, we employed features derived from the intensities and the histogram of the image, in addition to the Otsu's method \citep{otsu} for computing the threshold that separated target and non-target voxels. This alternative relies on the maximization of the between-class variance to separate the voxels belonging to the different classes. Figure \ref{fig:lung_segmentation} shows the results obtained by the segmentation scheme.

After lung segmentation, the resulting images were registered employing the Elastix software \citep{kasper}. This process consists on finding a coordinate transformation \begin{math} T(\mathbf{x}) \end{math} that modifies a moving image \begin{math} \mathbf{I}_{M}(\mathbf{x}) \end{math} to be aligned with a fixed image \begin{math} \mathbf{I}_{F}(\mathbf{x}) \end{math}. A transformation model \begin{math} T_{\mu}(\mathbf{x})\end{math}, with parameters \begin{math} \boldsymbol{\mu} \end{math}, can be formulated as an optimization problem in which a cost function \begin{math} C \end{math} is minimised with respect to \begin{math} \boldsymbol{\mu} \end{math}, as follows:

\begin{equation}
\label{eq:register}
\hat{\boldsymbol{\mu}} = \argmin_{\boldsymbol{\mu}} \hspace{0.3mm} C \hspace{0.3mm}  (T_{\mu}; I_{F}, I_{M})
\end{equation}

\begin{figure*}
\centering
\includegraphics[width=0.75\textwidth] {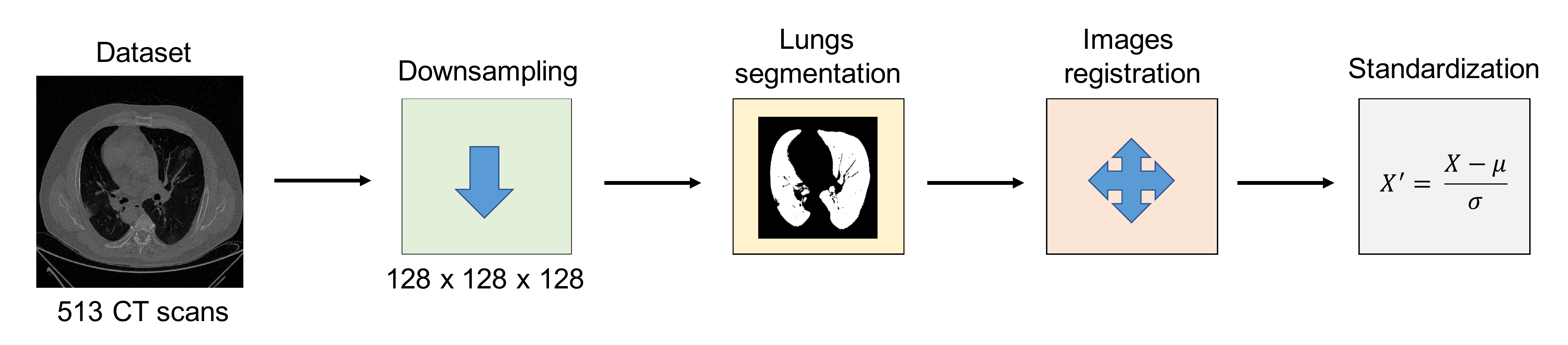}
\caption{Diagram of the preprocessing steps.}
\label{fig:preprocessing}
\end{figure*}

We employed the average of all the images in the database as the fixed image , whereas each individual image was iteratively selected as the moving one. We used the Mean Square Difference (MSD) as the cost function in order to evaluate the similarity between the fixed and the moving image. This function is defined as follows:

\begin{equation}
MSD(\mathbf{T}_{\boldsymbol{\mu}}; I_{F}, I_{M}) = \frac{1}{N} \sum_{\mathbf{x} \in \Omega_{F}}{(I_{F}(\mathbf{x}) -I_{M} (\mathbf{T}_{\boldsymbol{\mu}}(\mathbf{x})))^2}
\label{eq:cost}
\end{equation}

\noindent where \begin{math} \Omega_{F} \end{math} denotes the fixed image domain and \begin{math} N \end{math} the number of voxels \begin{math} \mathbf{x} \end{math} sampled from the domain of the fixed image. See \cite{Elastix} for a more detailed explanation. Finally, we performed an intensity normalization procedure for each individual image based on standardization. Each image was transformed such the resulting distribution had a zero mean and unit variance, as follows:

\begin{equation}
I' = \frac{I-\mu}{\sigma}
\label{eq:clahe3}
\end{equation}

\noindent where \begin{math} I \end{math} is the original image and \begin{math} I' \end{math} is the resulting one. Figure \ref{fig:preprocessing} shows a scheme of all the stages of the preprocessing.

\section{Methods}
\label{sec:methods}

\subsection{Kernel Principal Component Analysis}
\label{subsec:pca}
One of the main challenging problems in classification is related to the small sample size problem \citep{raudys}, which occurs when datasets are formed by high-dimensional data but with a small number of samples. Unfortunately, it is not uncommon to find large differences between the number of features and samples, so that finding a solution that alleviates this issue is crucial. Principal Component Analysis (PCA) is a multivariate approach which has been widely used to reduce the dimensionality of the data \citep{pca1,pca2,pca3}. This method attempts to find a linear subspace with a lower dimensionality than the original space. Given a set of \begin{math} N \end{math} samples \begin{math} \mathbf{x}_k \end{math}, \begin{math} \mathbf{x}_k  = [\mathbf{x}_{k1}, \ldots , \mathbf{x}_{kn} ] \in  \mathbb{R}^n\end{math}, the aim of PCA is to find the projection directions that maximize the variance of a subspace \citep{lopez}. This is equivalent to compute the eigenvalues from the covariance matrix. There are some occasions in which features can not be linearly extracted. In kernel PCA \citep{kernel_pca1,kernel_pca2}, vector \begin{math} \mathbf{x} \end{math} is projected from the input space, \begin{math}\mathbb{R}^n \end{math}, to a high-dimensional space,  \begin{math} \mathbb{R}^f \end{math}, by applying a non-linear mapping function \begin{math} \boldsymbol{\Phi}: \mathbb{R}^n \rightarrow \mathbb{R}^f, f>n  \end{math}. In the new feature space, \begin{math} \mathbb{R}^f \end{math}, the eigenvalue problem can be described as follows:

\begin{equation}
\label{eq:pca1}
C^{\Phi} \mathbf{w}^{\Phi} = \lambda \mathbf{w}^{\Phi}
\end{equation}

\noindent where \begin{math} C^{\Phi} \end{math} is a covariance matrix. All the solutions \begin{math} \mathbf{w}^\Phi \end{math} with \begin{math} \lambda \neq 0\end{math} are in the transformed space \begin{math} \Phi(\mathbf{x}_1,\ldots, \Phi(\mathbf{x}_N)) \end{math}, and there exist coefficients 
\begin{math} \alpha_i \end{math} such that:

\begin{equation}
\label{eq:pca2}
\mathbf{w}^\Phi = \sum_{i=1}^{N} \alpha_{i} \Phi(\mathbf{x}_i)
\end{equation}

Defining an \begin{math} N x N\end{math} matrix \begin{math} K \end{math} by 
\begin{equation}
\label{eq:pca3}
K_{ij} = k(\mathbf{x}_i,\mathbf{x}_j) = \Phi(\mathbf{x}_i) \cdot \Phi(\mathbf{x}_j)
\end{equation}

\noindent the PCA problem becomes:

\begin{equation}
\label{eq:pca4}
N \lambda K \boldsymbol{\alpha} = K^{2} \boldsymbol{\alpha} \equiv N\lambda \boldsymbol{\alpha} = K \boldsymbol{\alpha}
\end{equation}

\noindent where \begin{math} \boldsymbol {\alpha} \end{math} denotes a column vector with entries \begin{math} \alpha_{1} \ldots \alpha_{N}\end{math} \citep{kernel_pca2}.

A nonlinear version of PCA is obtained when using a nonlinear kernel such as the radial basis function (RBF), defined as follows:

\begin{equation}
\label{eq:pca5}
k(\mathbf{x}_{i},\mathbf{x}_{j}) = exp \Bigg(\frac{-0.5 \norm{\mathbf{x}_{i}-\mathbf{x}_{j}}^2}{\sigma^2} \Bigg)
\end{equation}

Finally, vectors in the high-dimensional feature space are projected into a lower dimensional spanned by the eigenvectors \begin{math} \mathbf{w}^\Phi \end{math}. Given a sample \begin{math} \mathbf{x} \end{math} whose projection is \begin{math} \Phi(\mathbf{x}) \end{math} in \begin{math} \mathbb{R}^f \end{math}, the projection of  \begin{math} \Phi(\mathbf{x})\end{math} onto the eigenvectors \begin{math} \mathbf{w}^\Phi \end{math} is the nonlinear principal components corresponding to \begin{math} \Phi \end{math}, as follows:

\begin{equation}
\label{eq:pca6}
\mathbf{w}^{\phi} \cdot  \Phi(\mathbf{x}) = \sum_{i=1} ^{N} \alpha_{i}(\Phi(\mathbf{x}_{i}) \Phi(\mathbf{x})) = \sum_{i=1}^{N} \alpha_{i}K(\mathbf{x}_i,\mathbf{x})
\end{equation}

\begin{math} \end{math}

\begin{figure*}
\centering
\includegraphics[width=0.85\textwidth] {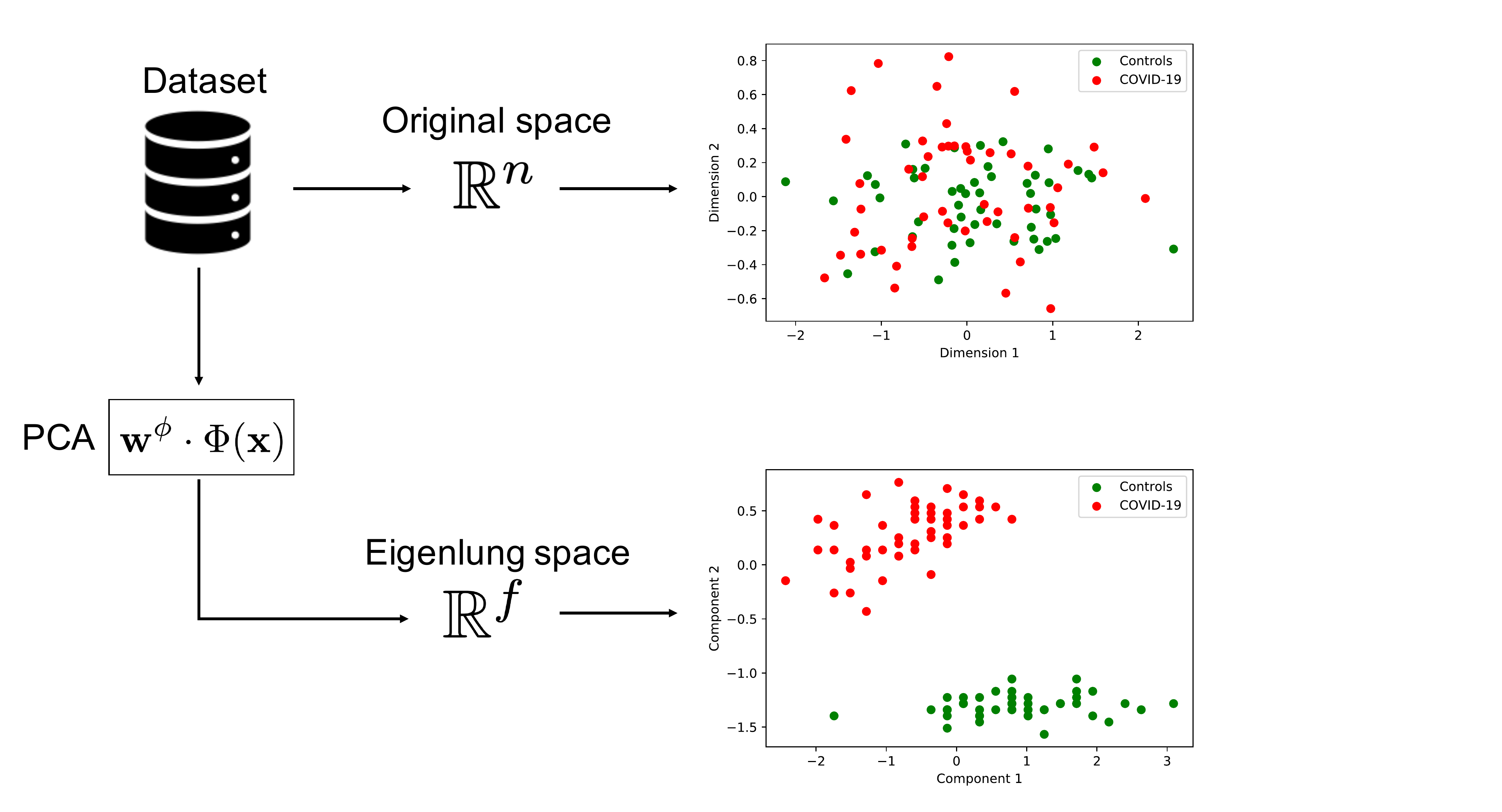}
\caption{Distribution of samples from the two classes before and after applying PCA. Projections onto the eigenlung space improve the separability between Controls and COVID-19 patients. The resulting eigenlungs that explain 90\% of the variance are then entered into the classifier.}
\label{fig:pca_esquema}
\end{figure*}

Derived from the \textit{eigenbrain} concept \citep{eigenbrain1,eigenbrain2,eigenbrain3} and given that sample \begin{math} \mathbf{x} \end{math} contains information from lungs, the \textit{eigenlungs} correspond to the dominant eigenvectors of the covariance matrix described in Equation \ref{eq:pca1}. The obtained \textit{eigenlungs} span a new subspace known as the \textit{eigenlung} space. From this space, we used the number of \textit{eigenlungs} that explained 90\% of the total variance, ranging from 20 to 30 \textit{eigenlungs} for the different experiments conducted. Figure \ref{fig:pca_esquema} depicts a visual description of the effect of PCA on data.

\subsection{Support Vector Machines}
\label{subsec:svm}
The resulting eigenlungs were then entered into an SVM classifier with RBF kernel \citep{cortes}, as follows:

\begin{equation}
\label{eq:svm1}
r_i = sign \Big( \sum_{i=1}^{N_{sv}} \alpha_i y_i K(\mathbf{x}_i, \mathbf{z}_j) +b\Big)
\end{equation}

\noindent where \begin{math} r_i \end{math} is the classification response for sample \begin{math} i\end{math}; \begin{math}N_{sv} \end{math} is the number of support vectors; \begin{math}\alpha_i \end{math} is the Lagrange multiplier; \begin{math} y_i \end{math} is the class membership of sample \begin{math} i \end{math}; \begin{math} K(\mathbf{x}_i, \mathbf{z}_j) \end{math} is the kernel function and \begin{math} b \end{math} is the bias parameter \citep{MORAIS201940}. Since classes were unbalanced (four times more pneumonia patients than controls), we incorporated the weights of the classes into the cost function of the SVM in order to assign to each sample a different relevance in the classification decision \citep{1555965,KinZen01}. This means that samples from the majority class had a lower influence in the penalty term than the ones from the minority class. 

The output of the classifier informs us about the class each sample belongs to. When classifying medical images it is particularly convenient to know not only if a patient suffers or not a disease but a degree of certainty of the prediction. However, standard SVMs do not provide any additional information of the predictions. \cite{Vapnik99thenature} proposed a method for mapping the outputs of SVMs to probabilities. This was based on the decomposition of the feature space into a direction orthogonal to the separating hyperplane and the rest of dimensions of the feature space. Despite its good performance, this approach requires a linear solution for every evaluation of the SVM. \cite{Platt99probabilisticoutputs} suggested an alternative based on training a logistic regression model on the classifier outputs in order to transform them into a probability distribution. The posterior probability can be defined as follows:

\begin{equation}
\label{eq:posterior1}
P(y=1|f) = \frac{1}{1+exp(Af+B)}
\end{equation}

\noindent The parameters \begin{math} A \end{math} and \begin{math} B \end{math} are fit using maximum likelihood estimation from a training set \begin{math} (f_i,y_i) \end{math}. Let \begin{math} (f_i,t_i) \end{math} a new training set, where \begin{math} t_i \end{math} are target probabilities defined as:

\begin{equation}
\label{eq:posterior2}
t_i = \frac{y_{i}+1}{2}
\end{equation}

\noindent Parameters can be estimated by minimizing the negative log likelihood of the training data, which is a cross-entropy error function:
\begin{equation}
\label{eq:posterior3}
min \Big[-\sum_{i}t_i \hspace{0.5mm} log(p_i)+(1-t_i) \hspace{0.5mm} log(1-p_i) \Big]
\end{equation}

\noindent where

\begin{equation}
\label{eq:posterior4}
p_i = \frac{1}{1+exp(Af_i+B)}
\end{equation}

We employed Equation \ref{eq:posterior1} for computing the probability that a test sample belonged to a specific class (control or covid patient). However, this measure itself doest not quantify the uncertainty of the prediction. To do so, we used a Bootstrap method following a random sampling with replacement scheme \citep{MR515681,WEHRENS200035}. This process consists on selecting part of the training set (80\% in our case), training the SVM classifier and computing the posterior probability of each test sample. Then, a new subset is randomly picked up from the training images and computed again the posterior probability of the test sample. This operation was repeated 500 times in order to build a distribution of probabilities. Finally, the uncertainty for a specific test sample \begin{math} k \end{math} was computed as the variance of the posterior probabilities, as follows:

\begin{equation}
\label{eq:posterior5}
u_{i} = \frac{\sum_{i=1}^{K} (x_i-\mu)^2}{K}
\end{equation}

\noindent where \begin{math} x_i \end{math} is the i-th element of the probabilities distribution \begin{math} x \end{math}, \begin{math} \mu  \end{math} is the mean of the distribution and  \begin{math} K \end{math} is the number of times that the Boostrapping process is repeated. Section \ref{subsec:ensemble} provides a detailed explanation about the use of uncertainty in our classification framework.

\subsection{Ensemble Classification}
\label{subsec:ensemble}

The development of a CAD (Computer-aided diagnosis) system for the detection of pneumonia relies on the assumption that patterns associated with this pathology are similar among patients. COVID-19 pneumonia usually presents GGO with a peripheral and subpleural distribution \citep{HANI2020263,opacities2}. However, the location and amount of these opacities may differ depending on the virulence of the disease. The presence of other pulmonary findings can also affect the reliability of the classification system, in addition to the artifacts during the acquisition of the images. To overcome these potential pitfalls, we divided each CCT scan into different cubic patches \citep{aortiz2016}. For each individual patch, kernel PCA was applied and the resulting components were then entered into the classifier. The number (and hence, size) of the patches were selected in order to match the potential size of the pneumonia patterns, guaranteeing to strike a balance between performance and computational cost. Finally, each individual classifier was then combined into a global one following an ensemble classification procedure. 

Majority voting has been widely used as a way of fusing the output of individual classifiers into a global decision \citep{chandra2021,zhou2020}. In this work, we computed the weight associated with each patch according to the uncertainty derived from the posterior probabilities obtained by the SVM classifier. If the variance of the probabilities of a classifier in a specific prediction was high, its contribution to the final ensemble would be low, and viceversa \citep{ensemble2012}. Defining \begin{math} u_{l}^{k}(\mathbf{y}) \end{math} as the uncertainty of the test sample \begin{math} \mathbf{y} \end{math} obtained from the \textit{k}-th classifier corresponding to the \textit{l}-th class, the empirical average of the \textit{l}-th weights (inverse of uncertainties) over the \textit{K} classifiers can be calculated as follows:

\begin{equation}
\label{eq:ensemble1}
E_{l}(\mathbf{y}) = \frac{\sum_{k=1}^{K}{\frac{1}{u_{l}^{k}(\mathbf{y})}}}{K}
\end{equation}

The class label of the test sample \begin{math} \mathbf{y} \end{math} is then assigned to the class with the maximum average weight as:

\begin{equation}
\label{eq:ensemble2}
Label(\mathbf{y}) = \argmax_{l} \hspace{0.5mm} E_{l}(\mathbf{y})
\end{equation}

\begin{figure*}
\centering
\includegraphics[width=0.75\textwidth]{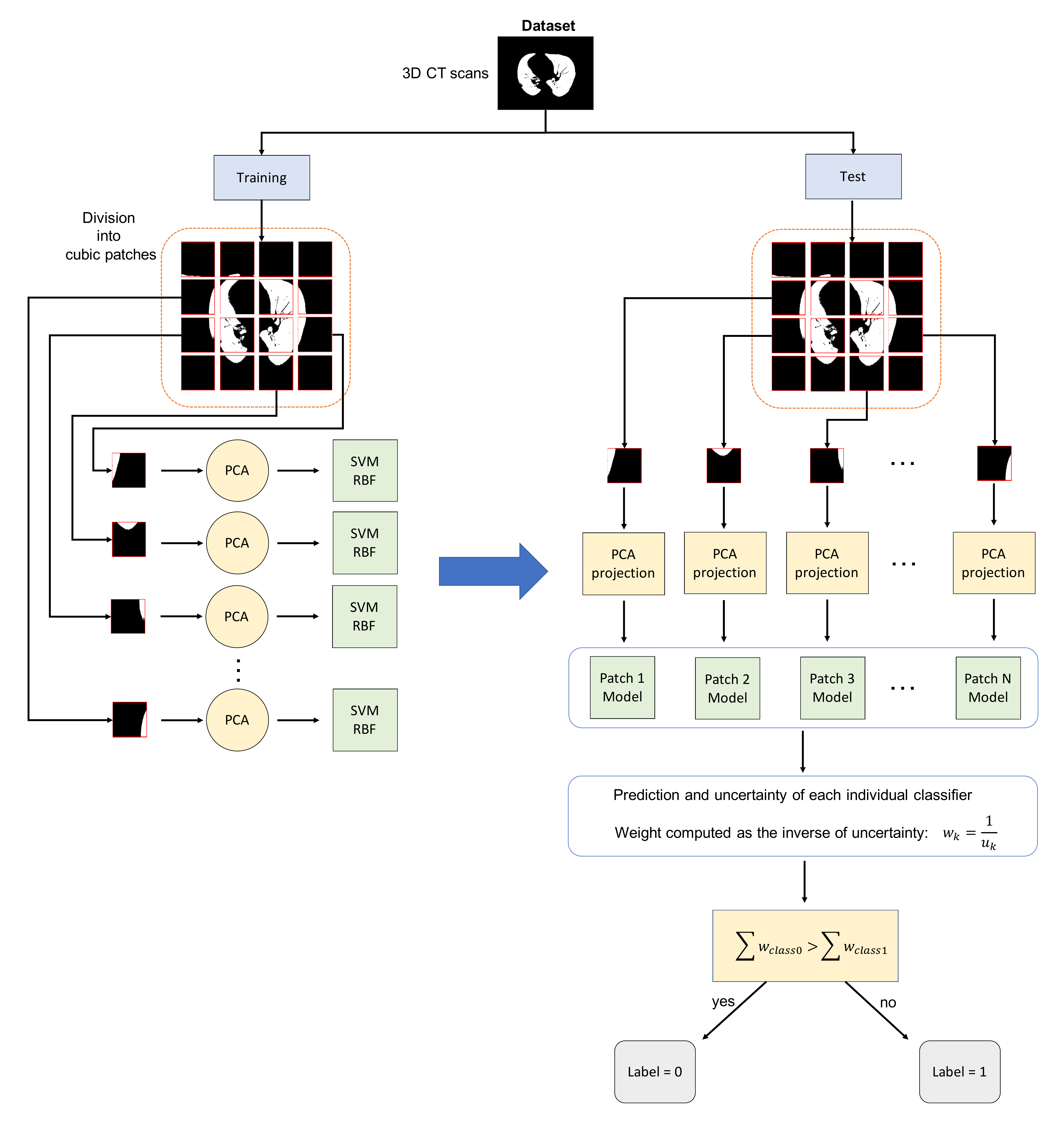}
\caption{Schema of the classification system proposed in this work.}
\label{fig:ensemble_scheme}
\end{figure*}

Figure \ref{fig:ensemble_scheme} shows a scheme of the ensemble classification system proposed.

\subsection{Performance Evaluation}
\label{subsec:performance}
We employed a Leave-One-Out cross-validation scheme to estimate the generalization ability of our method \citep{Kohavi95}. Specifically, from the total number of CCT scans (513), 512 were employed to train the model, whereas the remaining one was used for testing. This was performed within an iterative process in which all scans were used at some point as the test sample.The performance of the classification framework was evaluated in terms of different metrics derived from the confusion matrix: balanced accuracy, sensitivity, specificity, precision and F1-score, computed as follows:

\begin{align}
\label{eq:metrics}
\begin{split}
Bal \hspace{0.1cm} Acc = \frac{1}{2} \Big( \frac{TP}{P} + \frac{TN}{N}\Big)   \hspace{0.3cm} Sens = \frac {T_{P}}{T_{P}+F_{N}}  
\\
Spec = \frac {T_{N}}{T_{N}+F_{P}} \hspace{0.5cm} Prec = \frac {T_{P}}{T_{P}+F_{P}}
\\
F1-score = \frac{2 \times Prec \times Sens}{Prec + Sens}
\end{split}
\end{align}

\noindent where \begin{math}  T_{P} \end{math} is the number of pneumonia patients correctly classified (true positives), \begin{math} P \end{math} is the number of pneumonia patients, \begin{math} T_{N} \end{math} is the number of controls correctly classified (true negatives), \begin{math} N \end{math} is the number of controls, \begin{math} F_{P} \end{math} is the number of controls classified as pneumonia (false positives) and \begin{math} F_{N} \end{math} is the number of pneumonia patients classified as controls (false negatives), . The area under the ROC curve was also employed as an additional measure of the classification performance \citep{auc1,auc2}.

One crucial aspect is to evaluate the level of agreement of the different classifiers within the ensemble. To do so, we used a kappa-uncertainty diagram \citep{rodriguez2006,wang2019}. This measure relies on Cohen's kappa coefficient \citep{cohen1960}, a statistic that compares an observed accuracy with an accuracy obtained by chance, providing a measure of how closely instances classified by a classifier match the ground truth \citep{kappa_statistic2004}. Cohen's kappa can be mathematically defined as:

\begin{equation}
\label{eq:cohen}
k = \frac{p_A-p_E}{1-p_E}
\end{equation}

\noindent where \begin{math} p_A \end{math} is the observed relative agreeement between two annotators, and \begin{math} p_E\end{math} is the probability of agreement by chance. Although acceptable kappa statistic values vary on the context, the closer to 1, the better the classification. Section \ref{sec:results} summarizes the kappa scores obtained by the different members of the ensemble, in addition to how accuracies of individual classifiers and kappa values are related.

\section{Evaluation}
\label{sec:eval}

\begin{table*}[ht]
\caption{Performance of the ensemble classification proposed in this work for the different patch sizes evaluated.}
\label{table:results1}
\begin{tabular*}{\textwidth}{@{\extracolsep{\fill}}ccccccc}
 \hline
 
 Patch size & Bal Acc (\%) & Sens (\%) & Spec (\%) & Prec (\%) & AUC (\%) & F1-score (\%) \\
 \hline
 \rowcolor{LightGray}
 \multicolumn{7}{c}{Baseline approach: Voxels as Features} \\
 28 x 28 x 28 & 58.73 \textpm 2.43 & 67.2 \textpm 1.92 & 59.14 \textpm 1.46 & 60.02 \textpm 2.04 & 59.39 \textpm 1.35 & 61.12 \textpm 1.86 \\
 \rowcolor{LightGray}
 \multicolumn{7}{c}{Ensemble approach: Kernel PCA} \\
24 x 24 x 24  & 96.89 \textpm 1.43  & 100 & 84.45 \textpm 0.87 & 96.42 \textpm 1.01 & 91.34 \textpm 0.76 & 98.08 \textpm 0.55\\
28 x 28 x 28 & 97.86 \textpm 0.76  & 100 & 90.18 \textpm 0.86 & 96.98  \textpm 1.02 & 95.31 \textpm 0.54 & 98.75 \textpm 0.45  \\
32 x 32 x 32  & 97.27 \textpm 0.98  & 100 & 86.98 \textpm 1.24 & 96.72  \textpm 1.12 & 93.13 \textpm 1.32 & 98.33 \textpm 0.52\\
42 x 42 x 42 & 89.68 \textpm 1.64 & 100  & 81.04 \textpm 1.09 & 88.82 \textpm 1.49 & 89.36 \textpm 1.46 & 94.08 \textpm 0.87\\
48 x 48 x 48 & 85.5 \textpm 1.47 & 100 & 71.35 \textpm 2.01& 93.44  \textpm 0.76 & 85.52 \textpm 1.99 & 96.61 \textpm 0.79\\
56 x 56 x 56 & 84 \textpm 1.52 & 100 & 68.19  \textpm 1.87 & 92.81 \textpm 0.56 & 84 \textpm1.76 & 96.27 \textpm 0.73\\
64 x 64 x 64 & 88.37 \textpm 1.09 & 99.27 \textpm 0.41 & 78 \textpm 1.48 & 94.91 \textpm0.67 & 88.64 \textpm1.03  & 97.04 \textpm 0.82\\
 \hline
\end{tabular*}
\end{table*}

\subsection{Experimental setup} 
\label{subsec:setup}
In this work we propose a system to identify the patterns associated with pneumonia in CCT images that improves the automatic diagnosis of COVID-19. To do so, we define two experiments:

\begin{itemize}[leftmargin=*]
\item{\textbf{Experiment 1: Classification between controls and COVID-19 patients}. The aim is to detect the presence of pneumonia patterns in the different patches each CCT is divided into. The classification system was based on an ensemble of  RBF-SVM classifiers, whose parameters \begin{math} \gamma \end{math} and \begin{math} C \end{math} were optimized by using a grid search within a 5-Fold Cross-Validation scheme. The final values used were \begin{math} \gamma = 3 \end{math} and \begin{math} C = 1 \end{math}. Besides, patches are only evaluated if 20\% of their voxels contain lung regions.}

\item{\textbf{Experiment 2: Evaluation of the effect of the patch size} in the ensemble classification performance. A crucial aspect in the system proposed is the size of the cubic patches used for each individual classifier. A wide range of values were employed (from 24 to 64) in order to check the existence of an optimum size, which would be clearly associated with the common extension of pneumonia patterns associated with COVID-19. Moreover, we study the relationship between the patch size and the uncertainty measures derived from kappa scores in order to guide the election of a proper cubic region as a member of the ensemble.}

\item{\textbf{Experiment 3: Evaluation of the level of agreement of the different classifiers within the ensemble}. It is of great relevance to measure the performance of each individual classifier and compare it with the one obtained by the ensemble. Besides, we study how this relationship varies for different patch sizes.}

\end{itemize}

\section{Results}
\label{sec:results}

\begin{figure*}
\centering
\includegraphics[width=1\textwidth]{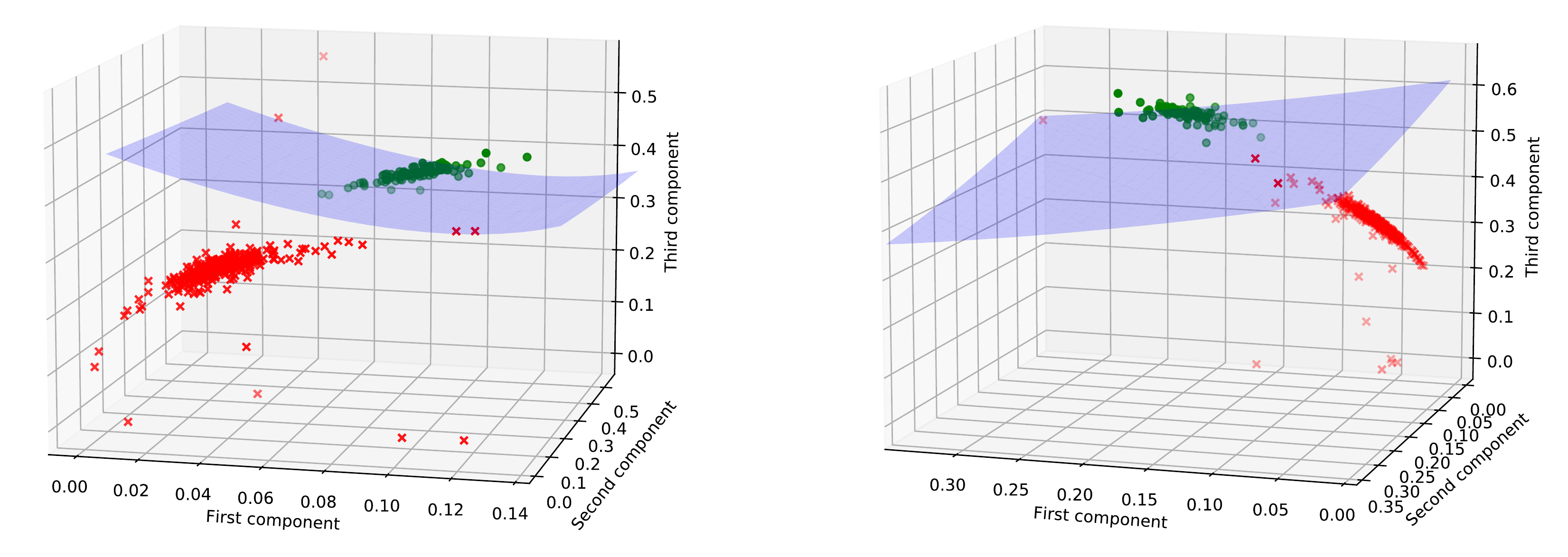}
\caption{Data image coefficients after polynomial kernel PCA projection and decision surfaces from the RBF-SVM classifier for two different patches. The number of eigenlungs needed for retaining 90\% of the variance ranges from 20 to 30. The hyperplane allows the separation between Controls and COVID-19 patients.}
\label{fig:eigenlungs}
\end{figure*}

\begin{figure*}
\centering
\includegraphics[width=0.6\textwidth]{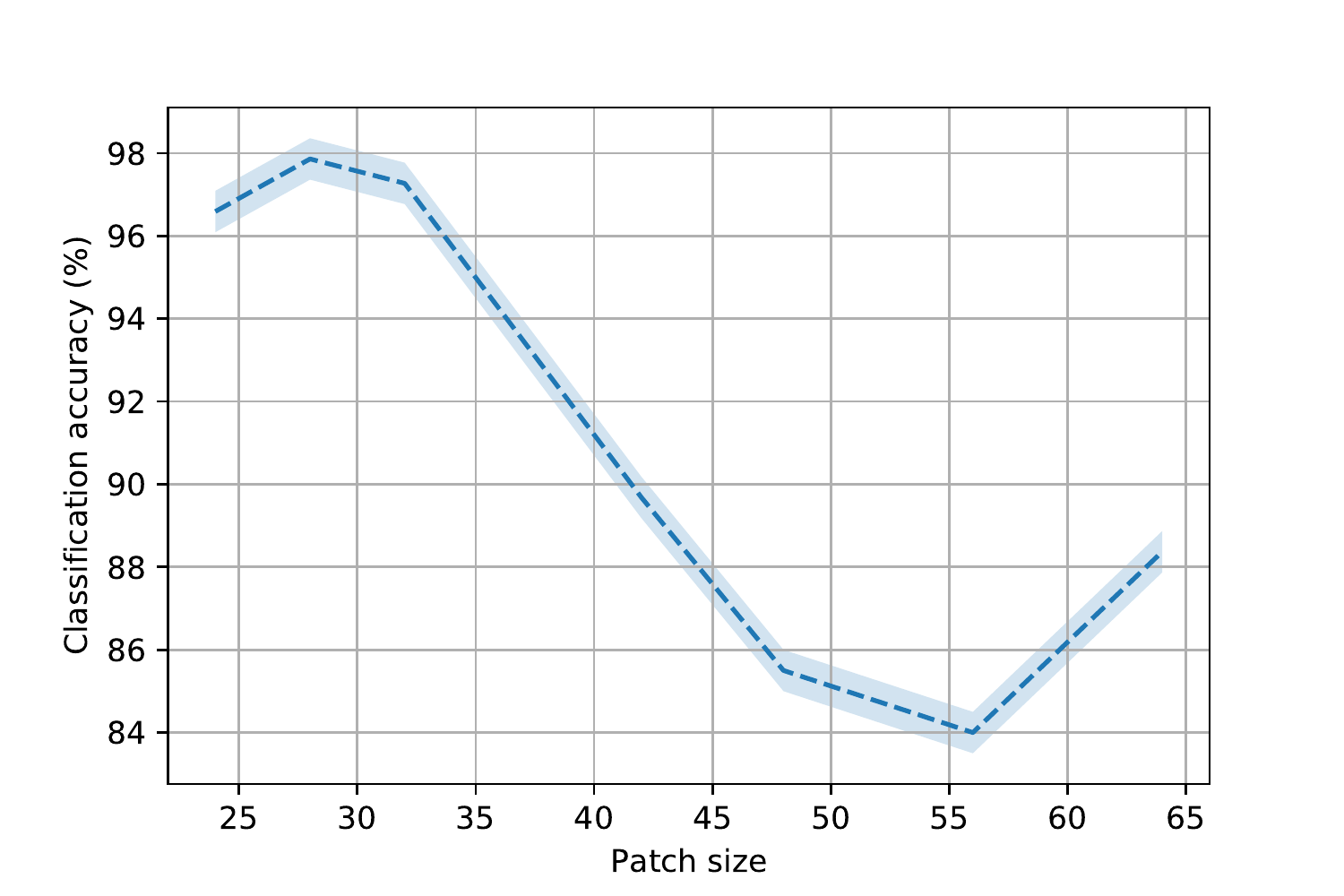}
\caption{Classification accuracies and their confidence bands obtained by the ensemble classifier for different patch sizes. The size is given in terms of the side of the cubic, so that a patch size = 24 refers to a cubic region of 24x24x24 voxels.}
\label{fig:acc_patch_size}
\end{figure*}

\begin{figure*}
\centering
\includegraphics[width=0.6\textwidth]{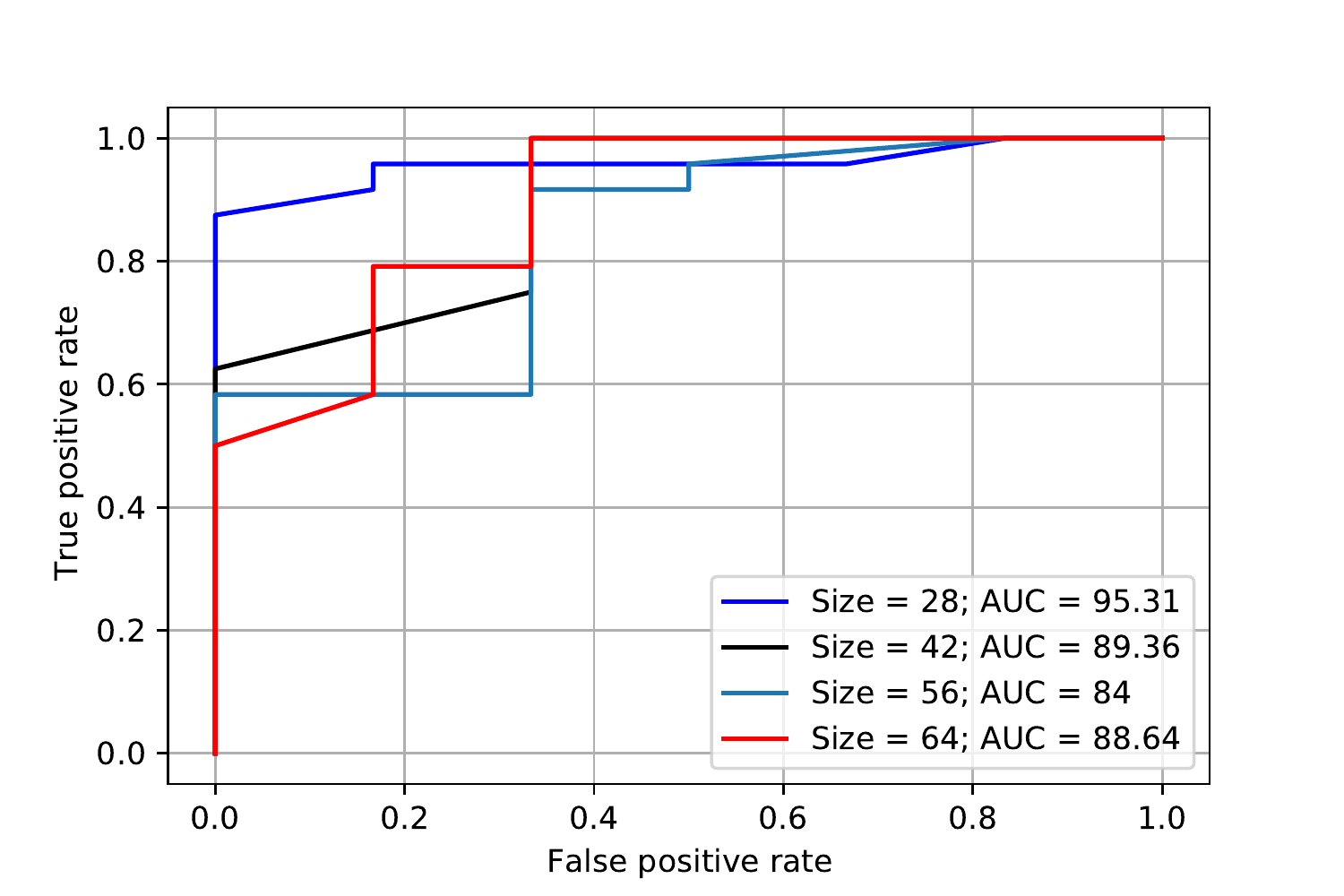}
\caption{ROC curves obtained by the ensemble approach for different patch sizes.}
\label{fig:roc_curves}
\end{figure*}

We first explore the performance of the ensemble classifier in terms of different measures, as summarized in Table \ref{table:results1}. We can see that the maximum accuracy obtained is 97.86\% when a cubic of 28x28x28 voxels is employed in each individual classifier. This manifests the high discrimination ability of the ensemble system. We also compared our system with a voxel-as-features (VAF) baseline method, in which all voxels of the patch are used as a feature vector. Results show that our approach highly outperformed the baseline, evidencing the need of applying feature extraction before classification. In fact, the projection provided by the eigenlungs allows the classifier to find an optimum solution for separating the two classes (see Figure \ref{fig:eigenlungs}). We also evaluate how performance differs according to the patch sizes of the base classifiers within the ensemble. Figure \ref{fig:acc_patch_size} provides a visual representation of the relationship between accuracy and patch size. Despite accuracies are lower than the one obtained by the cubic of 28x28x28 voxels, other patch sizes also lead to high performance, confirming that the system proposed can detect the presence of pneumonia even when the size of the patches is not ideal. However, results evidence that accuracy starts decreasing when too large patches are employed. A similar behavior occurs when referring to the area under the ROC curve, as Figure \ref{fig:roc_curves} shows. Patches that cover too wide regions can add confounds due to other anatomical structures unrelated to COVID-19 but with a similar appearance. Besides, the use of an exceeding size for the region covered by each base classifier can be detrimental for identifying the location of pneumonia. This can be especially relevant when the pulmonary affection of the patient is not severe.

\begin{figure*}
\centering
\includegraphics[width=0.6\textwidth]{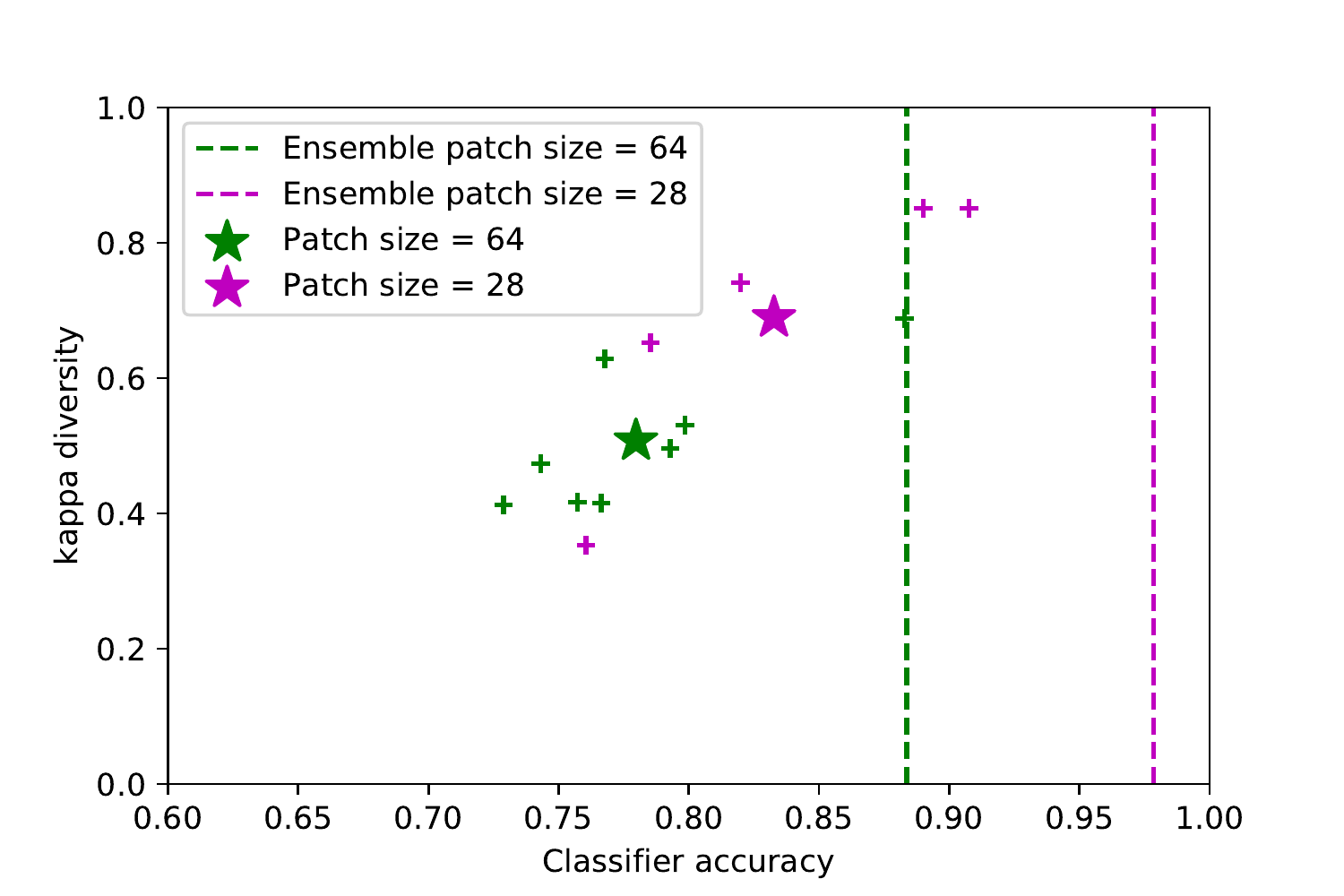}
\caption{Diversity-accuracy diagrams of the ensemble classifier for two different patch sizes. The x-axis represents the balanced accuracy obtained by each individual classifier and the resulting ensemble. The y-axis represents diversity of the classifiers evaluated by the kappa measure. Each marker represents the kappa-accuracy score obtained by by each individual classifier, whereas large stars represent the centroid of the resulting distribution. The dashed vertical lines represent the ensemble accuracies for the different patch sizes.}
\label{fig:kappa_acc}
\end{figure*}

We also use the kappa-accuracy diagram to evaluate the level of agreement between the classifier outputs. Figure \ref{fig:kappa_acc} shows these diagrams for two different patch sizes. The cloud points represent the kappa score-accuracy obtained by each individual classifier, whereas large stars represent the centroid of the resulting distribution. The most interesting aspects to highlight are the relationship between kappa score and accuracy of individual classifiers, in addition to the differential performance between each base classifier and the resulting ensemble. We can see that kappa diversity and accuracy are linearly dependent: a low kappa score leads to a low classification accuracy, and vice versa. Moreover, the accuracy of the ensemble classifier is much higher than the one obtained by individual members, supporting the suitability of this approach in this context. Although predictions of all base classifiers are used for taking the final decision, it is worth remembering that their contribution are weighted by the uncertainty of their predictions. If a classifier is much better than the others (in terms of accuracy and reliability), its decision will contribute much more than the rest. Figure \ref{fig:kappa_acc} shows an extreme case in which one base classifier overcomes by far the rest of the members of the ensemble. When the patch size was 64 (green line and markers), performance of the ensemble was exactly the same as performance of the best individual classifier, which is due to the large weight of this classifier compared to the rest. With reference to the patch size of 28, it shows a more desirable behavior: the final result is given by the combination of more than one member of the ensemble.

\section{Discussion}
\label{sec:discussion}

In this study, we proposed a classification system for the detection of pneumonia associated with COVID-19 from CCT scans. This approach relies on the use of probabilistic RBF-SVM classifier, which provides a measure about the reliability of the predictions in addition to the prediction itself. We partitioned CCT images into different patches and employed an ensemble classification framework so that individual predictions were combined into a global decision according to the reliability of each individual classifier. We evaluated the performance of this approach in terms of different measures and studied the influence of the patches size in the global performance and the relationship between individual and global decisions.

The high performance shown by the classification system proposed in this work led to an accurate tool for detecting the presence of pneumonia in CCT scans, in addition to spatially identifies where this affection is located. It is extremely important that a simple method like kernel PCA was able to extract the relevant information from each patch. There is a current tendency to use deep learning for the analysis of medical images, especially due to the large performance that these alternatives provide. However, there are some scenarios in which deep learning is not always the best choice for two main reasons. First, the use of convolutional blocks in 3D images requires a high amount of mathematical operations, leading to an excessive increase in the computational cost. Besides, the convergence and generalization ability of a convolutional neural network are highly influenced by the size of the dataset. Previous studies have performed dimensionality reduction of the input data by selecting only one slice from the 3D CCT. However, this  is not the optimal choice for several reasons. It is likely that using only one slice discards information that could be relevant for the classification process. Moreover, one of the main advantages of CCT over CXR is its high resolution and the tridimensional volumes that it provides, so that eliminating one of these features can be detrimental. Another important issue is that the process for selecting the slice that contains the pulmonary affection is not trivial. This is especially challenging when patients show an incipient pneumonia that is only located in a few lung regions. In this scenario, the selection of the slice would require the help of clinicians, eliminating the desired automaticity that our approach provides.

Another crucial aspect of our method is the way different members of the ensemble are combined. Unfortunately, COVID-19 often causes severe bilateral pneumonia, which means that the pathology is spread across large regions of the lungs. However, in first or intermediate stages of the disease only small pulmonary regions are damaged. This means that when a CCT image is divided into patches, most of them would be classified as controls since no pneumonia patterns would be found. When employing majority voting for the combination of individual classifiers within an ensemble, the final decision only depends on the number of patches that votes for a certain class. In the described scenario, all images from the first stages of the disease (where lung damage is scarce) would be labelled as control patients, invalidating its use as an accurate tool for the detection of pneumonia. However, since we weighted each member of the ensemble according to their uncertainty, the decisions of some members are more important than others. Patches that contain features similar to controls will lead to a high uncertainty. On the other hand, patches with lung lesions will be easily distinguished from controls, resulting in a low uncertainty and therefore, a high weight in the final decision. Thus, our system detects pneumonia for different grades of severity, from early stages to hyperinflammatory phase. This is extremely useful for an early diagnosis of the pathology and can help doctors to select the proper treatment that speeds up the recovery of the patient.

We have developed a complete system that is able to identify the patterns associated with pneumonia caused by COVID-19. It is worth highlighting the high performance obtained by our proposal: the accuracy and the AUC obtained were 97.86\% and 95.31\%, respectively. These results overcome other similar techniques in previous studies \citep{zhang2020_1,wang2020_ct,zhou2020,hemdan2020covidxnet,apostolopoulos2020}. There are some relevant aspects regarding our system to be mentioned. First, our system obtained excellent results while keeping a simple solution for the classification of tridimensional images. Second, the probabilistic nature of the classification scheme provides extremely useful information for clinicians. Our approach detects the presence (or not) of pneumonia and a measure of the uncertainty of its prediction, which can be converted in visual maps (patches with highest accuracy and lowest uncertainty) that help doctors to identify the pulmonary affection associated with COVID-19.

\section{Conclusion}
\label{sec:conclusion}
The collapse of the health system with the  outbreak of the COVID-19 pandemic has manifested the importance of finding a rapid, simple and accurate method for an early diagnosis of this disease. In this paper, we proposed an ensemble of probabilistic classifiers to detect pneumonia patterns associated with COVID-19. This scheme identified the lung regions from CCT scans and subdivided them into different patches. Then, features were extracted and entered into an RBF-SVM classifier. Predictions from different classifiers were weighted according to the uncertainty of their decisions, instead of using majority voting as most studies usually do. This allows to detect pneumonia even in its first stages, i.e. when affection is located in small pulmonary regions and most of them are similar to controls lungs. The large performance obtained (97.86\% of accuracy) and the simplicity of the system (no CNN employed) evidence the applicability of our proposal in a real-world environment. The combination of individual classifiers according to the uncertainty of their predictions provides an automatic tool that detects the presence of pneumonia, identifies its location and quantifies the reliability of the classification decision. Ours results pave the way for the development of new techniques that provide spatial information about where pulmonary affection is located. This can help clinicians not only in the patient's diagnosis but in the election of the proper treatment that rapidly alleviates symptoms associated with COVID-19.

\section*{Acknowledgments}\label{sec:Acknowledgments}
This work was partly supported by the Ministerio de Ciencia e Innovación (España)/ FEDER under the RTI2018-098913-B100 project, and by the Consejería de Economía, Innovación, Ciencia y Empleo (Junta de Andalucía) and FEDER under CV20-45250 and A-TIC-080-UGR18 projects. We want to thank HT M\'{e}dica for providing us with the CCT dataset employed in this work.

%
%
%
%

\vfill
\pagebreak

\bibliographystyle{elsarticle/elsarticle-harv}

\bibliography{biblio}

\end{document}